\documentclass[letterpaper,twocolumn,fleqn]{article} 
\usepackage{epsfig}
\usepackage{float}
\usepackage{subfig}
\usepackage{epstopdf}
\usepackage{color}
\usepackage{multirow}
\usepackage{hhline}
\usepackage{courier}
\usepackage{graphicx}
\usepackage{amsmath}
\usepackage{color}
\usepackage{ist}

\title{Multi-Reference Video Coding Using Stillness Detection}

\author{Di Chen$^{\star}$, Zoe Liu$^{\dagger}$, Yaowu Xu$^{\dagger}$, Fengqing Zhu$^{\star}$, Edward Delp$^{\star}$\\ $^{\dagger}$Google, Inc., 1600 Amphitheatre Parkway, Mountain View, CA, USA 94043.\\$^{\star}$School of Electrical and Computer Engineering, Purdue University, West Lafayette, Indiana, USA 47907.}

\date{} 

\begin{document} 

\maketitle 

\thispagestyle{empty} 


\begin{abstract}
Encoders of AOM/AV1 codec consider an input video sequence as succession of frames grouped in Golden-Frame (GF) groups. 
The coding structure of a GF group is fixed with a given GF group size.
In the current AOM/AV1 encoder, video frames are coded using a hierarchical, multilayer coding structure within one GF group. 
It has been observed that the use of multilayer coding structure may result in worse coding performance if the GF group presents consistent stillness across its frames.
This paper proposes a new approach that adaptively designs the Golden-Frame (GF) group coding structure through the use of stillness detection.
Our new approach hence develops an automatic stillness detection scheme using three metrics extracted from each GF group. 
It then differentiates those GF groups of stillness from other non-still GF groups and uses different GF coding structures accordingly. 
Experimental result demonstrates a consistent coding gain using the new approach.
\end{abstract}

\section{Introduction}
\label{sec:intro}
The AOM/AV1 codec \cite{AOM} is an open source, royalty-free video codec developed by a consortium of major technology companies called Alliance for Open Media (AOM) which is jointly founded by Google. It followed the VP9 codec \cite{VP9,VP92}, a video codec designed specifically for media on the web by Google WebM Project \cite{webm}. The AOM/AV1 codec introduced several new features and coding tools such as switchable loop-restoration \cite{ICIP1}, global and locally warped motion compensation \cite{ICIP2}, and variable block-size overlapped block motion compensation \cite{ICIP3}. 
The AOM/AV1 is expected to achieve generational improvement in coding efficiency over VP9. 

Current AOM/AV1 codec divides the source video frames into Golden-Frame (GF) groups. 
The length of each GF group, i.e. the GF group interval, may vary according to the video's spatial or temporal characteristics and other encoder configurations, such as the key frame interval at request for the sake of random access or error resilience. 
The coding structure of each GF group is based on their interval length and the selection of reference frames buffered for the coding of other frames. The coding structure determines the encoding order of each individual frame within one GF group.

\begin{figure}[h]
\centering
\subfloat[GF Group Coding Structure Using Multilayer]{
	\label{subfig:multi}
	\includegraphics[width=0.4\textwidth]{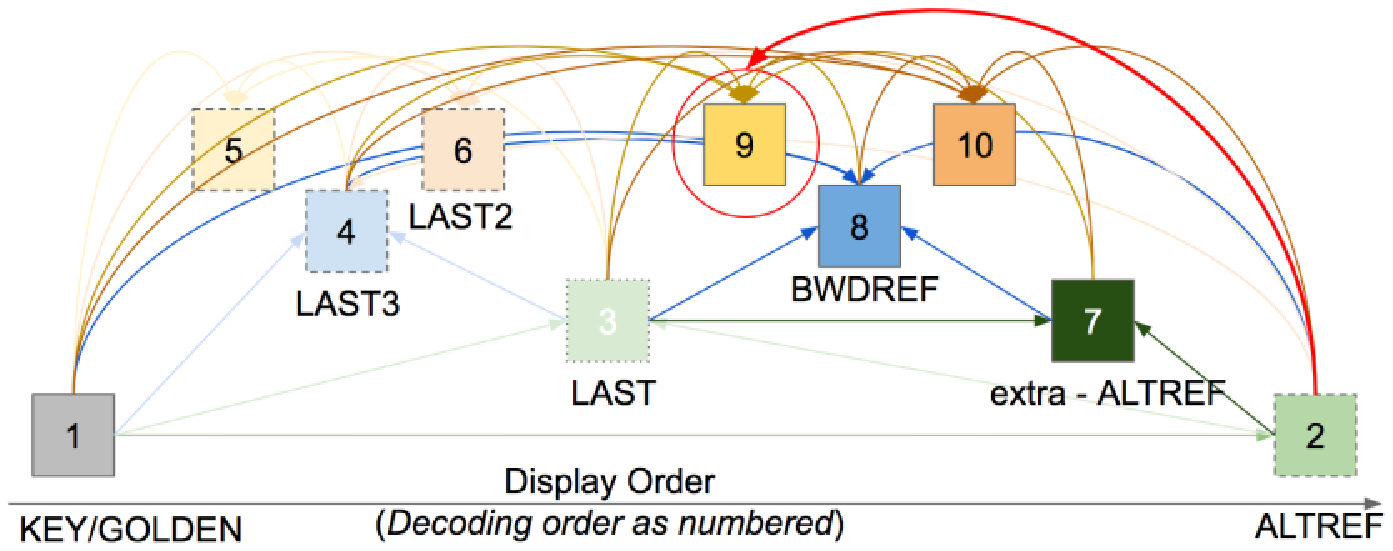} } 
 
\subfloat[GF Group Coding Structure Using One Layer]{
	\label{subfig:single}
	\includegraphics[width=0.4\textwidth]{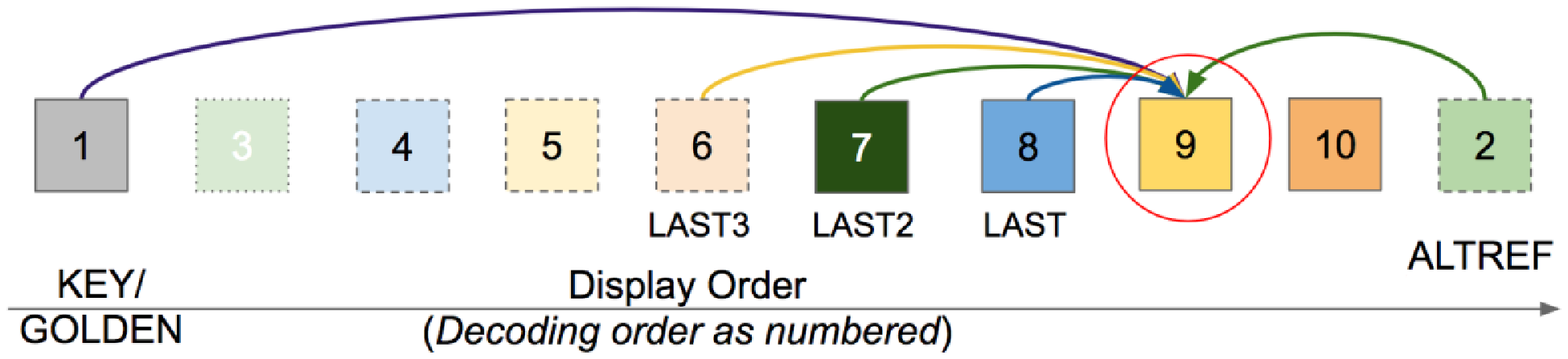} } 
\vspace*{5mm}
\caption{GF Group Coding Structures}
\label{structure}
 
\end{figure}

In the current implementation of the AOM/AV1 encoder, a GF group may have a length between 4 to 16 frames.
Various GF coding structures may be designed depending on the encoder'€™s decision on the construction of the reference frame buffer, as shown in Figure \ref{subfig:multi} and Figure \ref{subfig:single}. 
The \texttt{extra-ALTREF\_FRAME}s and the \texttt{BWDREF\_FRAME}s introduce hierarchical coding structure to the GF groups \cite{multiref}. The VP9 codec uses three references for motion compensation, namely \texttt{LAST\_FRAME}, \texttt{GOLDEN\_FRAME} and \texttt{ALTREF\_FRAME}. \texttt{GOLDEN\_FRAME} is the intra prediction frame.
\texttt{LAST\_FRAME} is the forward reference frame.
\texttt{ALTREF\_FRAME} is the backward reference frame selected from a distant future frame. It is the last frame of each GF group. 
A new coding tool is adopted by AV1 that extends the number of reference frames by adding \texttt{LAST2\_FRAME}, \texttt{LAST3\_FRAME}, \texttt{extra-ALTREF\_FRAME} and \texttt{BWDREF\_FRAME}. \texttt{LAST2\_FRAME} and \texttt{LAST3\_FRAME} are similar to \texttt{LAST\_FRAME}. \texttt{extra-ALTREF\_FRAME} and \texttt{BWDREF\_FRAME} are backward reference frames in a relatively shorter distance. 
The main difference is that \texttt{BWDREF\_FRAME} does not apply temporal filtering.
The hierarchical coding structure in Figure \ref{subfig:multi} may greatly improve the coding efficiency due to its multi-layer, multi-backward reference design. 

The current AOM/AV1 encoder uses the coding structure shown in Figure \ref{subfig:multi} for all the GF groups. 
However, a comparison of the compression performance with \texttt{extra-ALTREF\_FRAME} and \texttt{BWDREF\_FRAME} enabled and disabled showed that the coding efficiency for some test videos was actually worse when these two reference frames were enabled.
This means that the multilayer coding structure does not always have better coding efficiency for all the GF groups. One such example is the GF groups with stillness feature. 
In this paper, we propose a new approach that adaptively designs the Golden-Frame (GF) group coding structure through the use of stillness detection. A set of metrics are designed to determine whether the frames in a GF group is of little motion. 
Little work has been done that investigates the use of difference coding structures depending on video content. In \cite{Shih}, an adaptive video coding control scheme is proposed that suggests using more P- and B-frame while the temporal correlation among the frames in a group of pictures (GOP) are high. A method for using different GOP size based on video content is presented in \cite{Bruno}.  


\section{Method}

\subsection{GF Group Stillness}
A GF group may be constructed to contain consistent characteristics to differentiate itself from other GF groups. 
For instance, some GF group may present stillness across its successive frames, and other may present a zoom-in / zoom-out motion across the entire GF group. 
We examined the coding efficiency and the stillness feature of each GF group and found that 
when stillness is present in one GF group, the use of multilayer coding structure as shown in Figure \ref{subfig:multi} may produce worse coding performance, as opposed to that generated by the one layer structure in Figure \ref{subfig:single}.

\subsection{Automatic GF Group Stillness Detection}
\begin{figure}[!ht]
\centering
\subfloat[$zero\_motion\_accumulator$]{
	\includegraphics[width=0.4\textwidth]{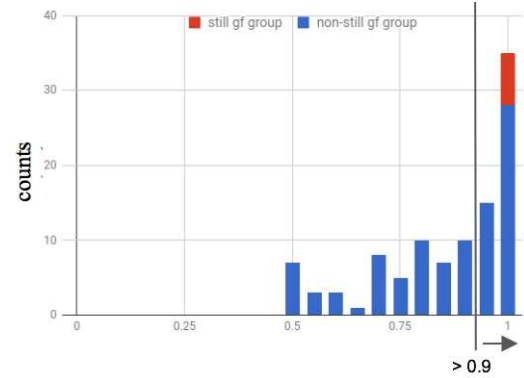} } 
 
\subfloat[$avg\_pixel\_error$]{
	\includegraphics[width=0.4\textwidth]{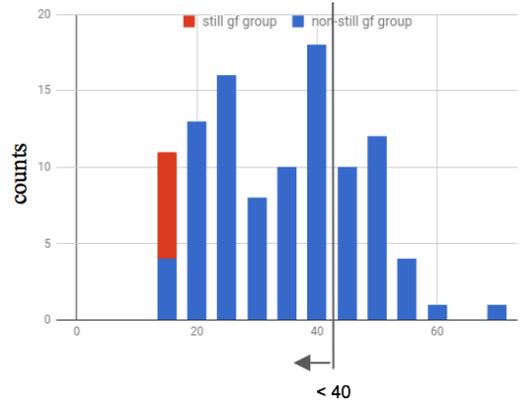} }\\ 
\subfloat[$avg\_error\_stdev$]{
	\includegraphics[width=0.4\textwidth]{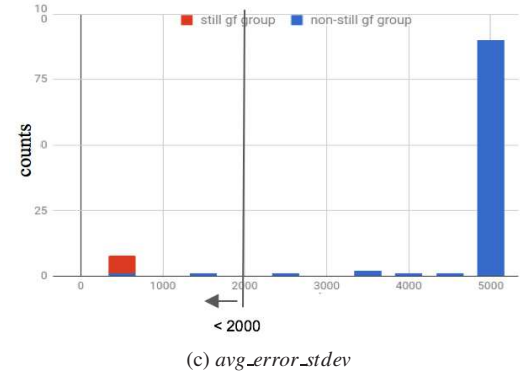} } 
\vspace*{5mm}
\caption{Thresholds for Metrics}
\label{figure:threshold}
\end{figure}

\begin{figure}[h]
  \includegraphics[width=0.9\columnwidth]{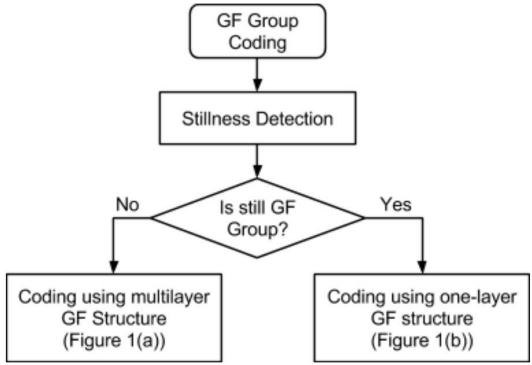}
  \caption{GF Group Coding With Stillness Detection}
  \label{figure:stillness}
\end{figure}

An automatic stillness detection of the GF groups is proposed in this paper which allows the GF groups to choose adaptively between two coding structure as shown in Figure \ref{subfig:multi} and Figure \ref{subfig:single}. 
Three metrics are extracted from the GF group during the first coding pass of AOM/AV1 to determine the GF group stillness.
The first coding pass of AOM/AV1 conducts a fast block matching with integer-pixel accuracy and use only one reference frame, the previous frame. Some motion vector and motion compensation information are collected during the first coding pass. 
Our proposed stillness detection method uses this information to extract three metrics as described below which requires small amount of computation. 
It then identifies the thresholds and derives the criteria to classify GF groups into two categories: GF groups of stillness and GF groups of non-stillness. 
The thresholds are obtained by collecting statistics of the three metrics from GF groups of eight low resolution (\textit{cif}) test videos. 
We manually labeled the stillness or non-stillness of the GF groups. 
Figure \ref{figure:threshold} shows the histograms and the thresholds of the three metrics.  
We intentionally included some test videos that contain GF groups of ``stillness-like" characteristics in the non-stillness class because they are more likely to be misclassified as GF group of stillness. 
The GF group with ``stillness-like" characteristics shows either very slow motion or static background with small moving objects.   
We obtained three criteria which are jointly applied to automatically detect stillness. 
Finally, the GF group is coded using the workflow given in Figure~\ref{figure:stillness}.

\textbf{Stillness Detection Metrics:}\\
\\
1. $\mathbf{zero\_motion\_acumulator}$: Minimum of the per-frame percentage of zero-motion inter blocks within one GF group:
{\setlength{\mathindent}{0cm}
\begin{equation}
zero\_motion\_accumulator = MIN (pcnt\_zero\_motion_{Fi}\ |\ Fi \in S)  
\end{equation}
where\\
$S = \{ Fi | i = 1, 2, ..., gf\_group\_interval \}$, the set of frames in the GF group\\
$gf\_group\_interval$: number of frames in the GF group\\
$pcnt\_zero\_motion$: percentage of the zero-motion inter blocks out of all the inter blocks\\
\\
2. $\mathbf{avg\_pixel\_error}$: Average of per-pixel sum of squared errors (SSE) within one GF group:
{\setlength{\mathindent}{0cm}
\begin{equation}
\begin{aligned}
& avg\_pixel\_error = \\
& MEAN(frame\_sse_{Fi} / number\ of\ pixels\ per\ frame\ |\ Fi \in S)
\end{aligned}
\end{equation}
where\\
$frame\_sse_{Fi}$ is the SSE of frame $Fi$\\
\\
3. $\mathbf{avg\_error\_stdev}$: First calculate the standard deviation of the block-wise SSEs for each frame, where block SSEs are obtained from zero-motion prediction; then obtain the mean value of the standard deviations of all the frames in one GF group:
\begin{equation}
avg\_error\_stdev = MEAN (STDEV_{Fi} (block\_sse_{(0,0)} )\ |\ Fi \in S)
\end{equation}
where\\
$block\_sse_{(0,0)}$ is the block-wise SSEs obtained from zero-motion prediction\\
$STDEV_{Fi}$ is the standard deviation of the block-wise SSEs of frame Fi\\

We use the above three metrics to differentiate those GF groups of stillness features from other GF groups, subject to the criteria in Table \ref{tab:fonts}.
\begin{table}[!ht]
	\caption{Table \ref{tab:fonts} Criteria for GF group stillness detection}
\label{tab:fonts}
\begin{center}       
\begin{tabular}{|p{0.45\columnwidth}|p{0.45\columnwidth}|} 
\hline
Stillness Detection&Stillness Detection\\ 
Metrics&Criteria\\
&(Identified as\\ 
&GF group of stillness) \\ \hline
$zero\_motion\_accumulator$& $>$0.9 \\ \hline
$avg\_pixel\_error$&$<$40 \\ \hline
$avg\_error\_stdev$&$<$2000 \\ \hline
\end{tabular}
\end{center}
\end{table} 

\subsection{Adaptive GF Group Structure Design}
Once a GF group is categorized as a GF group of stillness, no \texttt{extra-ALTREF\_FRAME} or \texttt{BWDREF\_FRAME} is used in the single layer coding structure as shown in Figure \ref{subfig:single}. 
The single layer coding structure still has multiple reference frames employed for the coding of one video frame. \texttt{LAST\_FRAME}, \texttt{LAST2\_FRAME}, \texttt{LAST3\_FRAME} and \texttt{GOLDEN\_FRAME} are used as forward prediction reference and \texttt{ALTREF\_FRAME} is used as backward prediction reference. 
If a GF group is categorized as non-still GF group, we will further leverage the use of \texttt{BWDREF\_FRAME} and \texttt{extra-ALTREF\_FRAME} to help improve the coding performance.

\section{Experimental Results}
We tested the proposed method using two standard video test sets with various resolutions and spatial/temporal characteristics, as shown in Table \ref{tab:result_aver}. More specifically, the set of \textit{lowres} includes 40 videos of \textit{cif} resolution, and the set of \textit{midres} includes 30 videos of 480p and 360p resolution. 
Each video is coded with a single \texttt{GOLDEN\_FRAME} and a set of target bitrates. 
For quality metrics we use the arithmetic average of the frame PSNR and SSIM \cite{ssim}. To compare RD curves obtained by the base AV1 codec and our proposed method, we use the BDRATE metric \cite{bdrate}. 
Experimental results demonstrated the advantage of the proposed approach. 
The Google test set of \textit{lowres} has two video clips that contain detected still GF groups (\textit{pamplet\_cif} and \textit{bowing\_cif}) and test set \textit{midres} has one (\textit{snow\_mnt}). 
As shown in Table \ref{tab:result_v}, by applying the proposed approach, the BDRATE 
of video clips that contains GF groups of stillness has decreased by approximately 1\%. The classification results of the proposed automatic stillness detector contains no misclassification case in the videos from these two test video sets. There are mainly two reasons that the single layer coding structure has better coding efficiency on the GF groups with stillness feature. 
One is that the multilayer coding structure in Figure \ref{subfig:multi} involves more candidate reference frames thus requires more motion information to be transmitted to the decoder. The other reason is that the multilayer coding structure uses an unbalanced bit allocation scheme which is not preferable for GF group of stillness in which the frames are very similar.    
 
\begin{table}[!h]
	\caption{Table \ref{tab:result_aver} BDRATE Reduction Using Proposed Method On Google Test Set}
\label{tab:result_aver}
\begin{center}       
\begin{tabular}{|p{0.2\columnwidth}|p{0.27\columnwidth}|p{0.25\columnwidth}|}
\hline
test set&BDRATE(PSNR)&BDRATE(SSIM)\\ \hline
test set of \textit{lowres}&-0.063&-0.045\\ \hline
test set of \textit{midres}&-0.026&-0.041\\ \hline
\end{tabular}
\end{center}
\end{table} 

\begin{table}[!h]
	\caption{Table \ref{tab:result_v} BDRATE Reduction Using Proposed Method On Video Clips Contain GF Group Of Stillness}
\label{tab:result_v}
\begin{center}       
\begin{tabular}{|p{0.2\columnwidth}|p{0.27\columnwidth}|p{0.26\columnwidth}|}
\hline
video clip&BDRATE(PSNR)&BDRATE(SSIM)\\ \hline
\textit{pamplet\_cif}&-1.395&-1.076\\ \hline
\textit{bowing\_cif}&-1.118&-0.735\\ \hline
\textit{snow\_mnt}&-0.767&-1.235\\ \hline
\end{tabular}
\end{center}
\end{table} 

\section{Conclusion and Future Work}
We proposed an automatic GF group stillness feature detection method. Each GF groups is classified into still GF group and non-still GF group based on three metrics and the encoder adaptively chooses the coding structure based on optimized coding efficiency. Experimental results showed coding gain for videos containing still GF group.
We also observed that GF groups containing other features, such as fast zoom-out and high motion, may also benefit from the single layer coding structure. 

\nocite{webm}
\nocite{VP10}
\nocite{Bref}
\nocite{VP8}
{\small
\bibliographystyle{ieee}
\bibliography{egbib}
}
\vfill\pagebreak

\begin{biography}

Di Chen is a PhD candidate in Video and Image Processing Laboratory (VIPER) at Purdue University, West Lafayette. Her research focuses on video analysis and compression. Currently, she is working on texture segmentation based video compression using convolutional neural networks. 
\\

Zoe Liu received her Ph.D. in ECE from Purdue University. She once worked with corporate research institutes, prototyping mobile video conferencing platforms at Bell Labs (Lucent) and Nokia Research Center. Zoe then devoted her effort to the design and development of FaceTime at Apple, Tango Video Call with the startup of TangoMe, and Google Hangouts. She is currently working in the WebM team of Google.
\\

Dr. Yaowu Xu is currently the Tech Lead Manager of the video coding research team at Google.  Dr. Xu's education background includes the BS degree in Physics, the MS and PhD degree in Nuclear Engineering from Tsinghua University at Beijing, China. He also holds the MS and PhD degree in Electrical and Computer Engineering from University of Rochester. His current research focuses on advanced algorithms for digital video compression.
\\

Fengqing Zhu is an Assistant Professor of Electrical and
Computer Engineering at Purdue University, West Lafayette, IN.
Dr. Zhu received her Ph.D. in Electrical and Computer Engineering
from Purdue University in 2011. Prior to joining Purdue
in 2015, she was a Staff Researcher at Huawei Technologies
(USA), where she received a Huawei Certification of Recognition
for Core Technology Contribution in 2012. Her research interests
include image processing and analysis, video compression,
computer vision and computational photography.
\\

Edward J. Delp was born in Cincinnati, Ohio. He is currently The Charles
William Harrison Distinguished Professor of Electrical and Computer
Engineering and Professor of Biomedical Engineering at Purdue
University. His research interests include image and video processing,
image analysis, computer vision, image and video compression, multimedia
security, medical imaging, multimedia systems, communication and
information theory. Dr. Delp is a Life Fellow of the IEEE, a Fellow of
the SPIE, a Fellow of IS\&T, and a Fellow of the American Institute of
Medical and Biological Engineering.

\end{biography}

\end{document}